\documentclass[11pt]{article}
\usepackage{geometry}     % See geometry.pdf to learn the layout options. There are lots.
\pdfoutput=1 %for pdflatex submission to arxiv
\usepackage{graphicx}
\usepackage{amsmath,amssymb}  % Better maths support & more symbols
\usepackage{bm,bbm}  % Define \bm{} to use bold math fonts
\usepackage{slashed}
\newcommand\beq{\begin{equation}}
\newcommand\eeq{\end{equation}}

\newcommand\diag{\text{diag}}
\newcommand\hc{\text{h.c.}}
\newcommand\LLH{$\text{L}^2\text{H}$}
\newcommand\PD{{\phantom{\dagger}}}
\newcommand\sfrac[2]{{\textstyle{\frac{#1}{#2}}}}

\title{Hidden fine tuning in the quark sector of little higgs models}
\author{Benjam\'\i{}n Grinstein\thanks{bgrinstein@ucsd.edu} \, Randall
  Kelley\thanks{randallkelley@physics.ucsd.edu} and Patipan Uttayarat\thanks{puttayarat@physics.ucsd.edu}\\
University of California, San Diego; \\ Dept. of Physics, La Jolla, CA 92093-0315}
\date{\today \vskip-2in \hfill UCSD PTH 09-01\vskip2in}            % Activate to display a given date or no date

\begin{document}
\maketitle
\begin{abstract}
  In little higgs models a collective symmetry prevents the higgs from
  acquiring a quadratically divergent mass at one loop. By considering
  first the littlest higgs model we show that this requires a fine
  tuning: the couplings in the model introduced to give the top quark
  a mass do not naturally respect the collective symmetry. We show the
  problem is generic: it arises from the fact that the would be
  collective symmetry of any one top quark mass term is broken by
  gauge interactions.
\end{abstract}
\newpage
\section{Introduction}
%\subsection{}

Little Higgs (LH) models offer an alternative to the standard model in
which no fundamental scalars need be introduced (for reviews see
\cite{Schmaltz:2005ky}).  Generally, in LH models the Higgs is a composite
particle, bound by interactions that become strong at a scale
$\Lambda$. The mass of the Higgs is much less than $\Lambda$ as
the Higgs is a pseudo-Goldstone boson (PGB) of broken
global symmetries in the theory of the new strong interaction.

The global ``flavor'' symmetry $G_f$ of these models has a subgroup
$G_w$ that is weakly gauged. In the absence of this weak gauge force,
the flavor symmetry is broken spontaneously to a subgroup $H$ due to
hyper-strong interactions at the scale $\Lambda$. As a result, there
are massless Goldstone bosons that are coordinates on the $G_f/H$
coset space. Since the weakly gauged $G_w$ force breaks the flavor
symmetry explicitly, including its effects leads to some of the
Goldstone bosons (the would-be Goldstone bosons)  being eaten by the
Higgs mechanism and the rest becoming PGBs
acquiring small masses of order $\Lambda$ times a small symmetry
breaking parameter, the weak gauge coupling constant.  The Higgs is
the lightest PGB in LH models, and its mass is naturally much less
than $\Lambda$ (and the other PGBs): due to the
collective symmetry breaking mechanism its mass arises only at two loops.

Additional interactions must be included in LH models to account for
quark and lepton masses. At low energies they  reproduce
the Yukawa couplings of the standard model. Since these interactions
also break the flavor symmetry, they contribute to the masses of the 
PGBs. In order to ensure that the Higgs remains much lighter than
$\Lambda$, the quark interactions are designed to implement the
collective symmetry breaking mechanism again. 

In this paper we observe that the quark interactions introduced in the
littlest higgs\cite{ArkaniHamed:2002qy} model ($\text{L}^2\text{H}$)
and variants are secretly fine tuned to ensure the Higgs remains
light.  In the \LLH\ we will show that radiative effects force the
coupling of the top-quark `triplet' to split into two terms,
disrupting the collective symmetry mechanism. We will argue that
therefore combining the two terms into a single one was the result of
an implicit fine tuning. We then show that the generic coupling of
the top-quark and partners, as allowed by symmetry, induces
unsuppressed, order $\Lambda$  Higgs masses. 

In other words, the top-quark couplings are usually taken to implement
the collective symmetry mechanism. But there is no underlying symmetry
that enforces this choice, which must therefore result only from fine
tuning. This defeats the purpose of introducing the model in the first
place. We present some details on the original \LLH\
model, and then proceed to show that  the problem is generic. 

Cannot one avoid this problem by gauging the collective symmetry?
After all, if the symmetry is gauged then the restricted form of the
quark coupling is a result of the gauge symmetry.  For example one may
construct a model based on $G_f/H=U(7)/O(7)$ with $G_w=SU(3)\times
SU(2)\times U(1)^3$.  The vacuum aligns\cite{Preskill:1980mz} so that
$G_w$ breaks to the electroweak subgroup $SU(2)\times U(1)$ at the
scale $\Lambda$, and the spectrum has a light Higgs doublet plus many
heavier PGBs. Could the gauged $SU(3)$ now play the role of the
collective symmetry for the top quark mass? The problem with this is
that the gauge symmetry is broken and the would be higgs is
eaten. This model is higgsless. This is also generic: the collective
symmetry must act nonlinearly on the higgs, and therefore it must be
broken. Gauging it eats away the higgs.

In Sec.~\ref{sec:llh} we review and explain the problem in the \LLH\
model. The  \LLH\ itself is phenomenologically disfavoured
\cite{Han:2003wu} by EWPD, and it is for this reason that
alternatives, like models with custodial symmetry\cite{Chang:2003un}
or with T-parity\cite{Cheng:2003ju}, have been introduced. Rather than
investigating these models individually we show in
Sec.~\ref{sec:no-go} that the problem is generic. We first give a very
explicit proof for models with $SU(N)/SO(N)$ (and $SU(N)/Sp(N)$) vacuum
manifold. We then generalize, which does not require much additional
work. A brief recap is in Sec.~\ref{sec:conc}.

\section{Top-quark coupling fine tuning in the Littlest Higgs Model}
\label{sec:llh}
\subsection{Model Review}
To establish notation we briefly review elements of the \LLH \,
\cite{ArkaniHamed:2002qy}. It has $G_f=SU(5)$, $H=SO(5)$ and
$G_w=\prod_{i=1,2}SU(2)_i\times U(1)_i$.  Symmetry breaking $SU(5)\to
SO(5)$ is characterized by the Goldstone boson decay constant $f$. The
embedding of $G_w$ in $G_f$ is fixed by taking the generators of
$SU(2)_1$ and $SU(2)_2$ to be
\begin{equation}
   Q^a_1=  
  \begin{pmatrix}
    \frac{1}{2}\tau^a&0_{2\times3}\\0_{3\times2}&0_{3\times3}
  \end{pmatrix}
  \qquad\text{and}\qquad
  Q^a_2=
  \begin{pmatrix}
    0_{3\times3}&0_{3\times2}\\0_{2\times3}&-\frac{1}{2}\tau^{a*}
  \end{pmatrix}
\end{equation}
and the generators of the $U(1)_1$ and $U(1)_2$ 
\begin{equation}
  Y_1=\frac1{10}\diag(3,3,-2,-2,-2)\qquad\text{and}\qquad
  Y_2=\frac1{10}\diag(2,2,2,-3,-3). 
\end{equation}

The vacuum manifold is characterized by a unitary, symmetric
$5\times5$ matrix $\Sigma$. We denote by $g_{i}$ ($g'_{i}$) the gauge
couplings associated with $SU(2)_i$ ($U(1)_i$). If one sets
$g_{1}=g_1'=0$ the model has an exact global $SU(3)$ symmetry (acting
on the upper $3 \times 3$ block of $\Sigma$), while for $g_{2}=g_2'=0$ it
has a different exact global $SU(3)$ symmetry (acting on the lower $3
\times 3$ block). 
Either of these exact global $SU(3)$ would-be
symmetries guarantee the Higgs remains exactly massless. Hence, the
Higgs mass should vanish for either $g_{1}=g_1'=0$ or
$g_{2}=g_2'=0$. The perturbative quadratically divergent correction to
the Higgs mass must be polynomial in the couplings and can involve
only one of the couplings at one loop order. Hence it must vanish at
one loop. This is the collective symmetry mechanism that ensures the
absence of 1-loop quadratic divergences in the higgs mass.

It is standard to introduce the top quark so that the collective
symmetry argument still applies. The third generation doublet $q_L$ is
a doublet under $SU(2)_1$ and a singlet under $SU(2)_2$.  Introduce
additional $SU(2)_1\times SU(2)_2$-singlet spinor fields: $q_R$, $u_L$
and $u_R$. The third generation right handed singlet is a linear
combination of $u_R$ and $q_R$.  The charges of these under
$U(1)\times U(1)$ are listed below, in \eqref{eq:Y1Y2}.  Their
couplings are taken to be
\begin{equation}
  \label{eq:Ltop}
  {\cal L}_{\text{top}}=
  -\frac12 \, \lambda_1 \, f \, \bar  \chi_{Li}^{\phantom{\dagger}}\, 
  \epsilon^{ijk} \, \epsilon^{xy} \,  \Sigma_{jx} \, \Sigma_{ky} 
  \ q_R^{\phantom{\dagger}} 
  - \lambda_2 \, f \, \bar u_L^{\phantom{\dagger}}\, u_R^{\phantom{\dagger}} +\hc
\end{equation}
where the indexes $i,j,k$ run over 1,2,3, the indexes $x,y$ over $4,5$
and the triplet $\chi_L^{\phantom{\dagger}}$ is 
\begin{equation}
  \label{eq:triplet}
  \chi_L^{\phantom{\dagger}}=
  \begin{pmatrix}
    i\tau^2 q_L^{\phantom{\dagger}}\\u_L^{\phantom{\dagger}}
  \end{pmatrix}.
\end{equation}
The collective symmetry argument now runs as follows. If $\lambda_2=0$
then ${\cal L }_{\text{top}}$ in \eqref{eq:Ltop} is constructed so
that it exhibits an explicit global $SU(3)$ symmetry, a subgroup of
$G_f=SU(5)$. Under this, the fields $\chi_L$ in \eqref{eq:triplet} and
$\Sigma_{ix}$ transform as triplets (on $i=1,2,3$). Since this
would-be exact global symmetry is spontaneously broken it guarantees
that the Higgs field remains an exactly massless Goldstone
boson. Similarly, if $\lambda_1=0$ then there is no coupling of the
quarks to the Goldstone bosons, which therefore remain
massless. Hence, the mass term must vanish as either $\lambda_1$ or
$\lambda_2$ are set to zero, and since the quadratic divergence is
polynomial in the couplings, it can only arise at two loops.

The gauge and top-quark interactions generate an effective,
Coleman-Weinberg potential which determines the vacuum orientation. If
the gauge couplings are strong enough\cite{Grinstein:2008kt},
\begin{equation}
  \label{eq:bound1}
  g_1^{\prime2}+g_1^2 >\frac{2N_c}{3\pi^2c} \, \lambda_1^2 \, \lambda_2^2 \, 
  \left[\ln\left(\frac{\Lambda^2}{(\lambda_1^2+\lambda_2^2)f^2}\right)+\frac{\hat
      c'}2\right].
\end{equation}
where $c$ and $\hat c'$ are unknown dynamical constants of order unity, the
vacuum alignment is
\begin{equation}
  \label{eq:Sigma0}
  \Sigma_{ew}=\begin{pmatrix} 0&0& \mathbf{1}_{2\times2}\\ 0 & 1& 0\\ \mathbf{1}_{2\times2} &0 &0\end{pmatrix}.
\end{equation}
leading to the gauge-symmetry breaking into the electroweak subgroup,
$\prod_{i=1,2}SU(2)_i\times U(1)_i\to SU(2)\times U(1) $.

\subsection{The Hidden Fine Tuning}
As we just saw, the top quark Lagrangian ${\cal L }_{\text{top}}$ in
\eqref{eq:Ltop} is constructed so that it exhibits an explicit global
$SU(3)$ symmetry.  However, this is a symmetry of the Lagrangian only
for $\lambda_2=g_1=g_1'=0$.

There is in fact no symmetry reason for the fields in $\chi_L$ to
combine into a triplet. Given that the effective Lagrangian is
restricted only by the non-linear realization of the symmetry (by
parametrizing $G_f/H$) and by the requirement of explicit gauge
invariance under $G_w$, the coupling in \eqref{eq:Ltop} is more
generally of the form
\begin{equation}
  \label{eq:Ltop-split}
  {\cal L}_{\text{top}}=-\lambda_1f\bar q_L^{\PD i}\epsilon^{xy}
  \Sigma_{ix}\Sigma_{3y}q_R^\PD
  -\frac12\lambda'_1f\bar u_{L}^\PD\epsilon^{3jk}\epsilon^{xy}
  \Sigma_{jx}\Sigma_{ky}q_R^\PD
  -\lambda_2f\bar u_L^\PD u_R^\PD +\hc
\end{equation}
Only when $\lambda_1'=\lambda_1$ (and $\lambda_2=g_1=g_1'=0$) do we
recover the global $SU(3)$ symmetry of the collective symmetry
mechanism. The main observation of this work is that the relation
$\lambda_1'=\lambda_1$, assumed throughout the little higgs literature,
is unnatural. We refer to this as the hidden fine tuning problem.

Although $\lambda'_1=\lambda_1$ is natural in the absence of the gauge
interactions, these are already present in the UV completion. Below
we comment in slightly more detail on how radiative effects explicitly
introduce $SU(3)$ breaking into the Yukawa couplings.

It should be evident that for $\lambda_1'\ne\lambda_1$ the collective
symmetry argument is spoiled. A straightforward computation gives a
quadratically divergent correction to the higgs mass,
\begin{equation}
  \label{eq:oops}
  \delta m_h^2
  =\frac{12}{16\pi^2}(\lambda^2_1-\lambda^{\prime2}_1)\Lambda^2
\end{equation}
where $\Lambda$ is a UV cut-off. The severity of the fine tuning can
now be explored. If we insist that the Higgs mass should be naturally
of order of $100$~GeV, while $\Lambda\sim 10$~TeV, then, not
surprisingly, 
$\lambda_1'-\lambda_1\lesssim (4\pi m_h/\Lambda)^2\sim1\%$.

The Lagrangian in \eqref{eq:Ltop-split} is not the most general one
consistent with symmetries to lowest order in the chiral
expansion. If $SU(3)$ were a good symmetry one could add to the
Lagrangian a term of the form
\begin{equation}
\label{eq:new-top-term}
  \bar \chi^\PD_{Li} \epsilon_{jkl}\epsilon_{xy}
  (\Sigma^*)^{ij}(\Sigma^*)^{kx}(\Sigma^*)^{ly}q_R^\PD
\end{equation}
One can also freely replace $q_R\leftrightarrow u_R$ in
Eqs.~\eqref{eq:Ltop} and~\eqref{eq:new-top-term}, and then, of course,
split each $SU(3)$ invariant term into a sum of $SU(2)\times U(1)$
invariant terms. There is no reason a priori why these terms should be
ignored, but they are not dangerous. In fact, they are
inevitable, as they are generated radiatively, many of them already at
one loop \cite{us-in-prep}.

\subsection{Radiatively induced $\lambda_1'\ne\lambda_1$}
Imposing $\lambda_1'-\lambda_1=0$ is not only a fine tuning, it  is
unnatural. Since the symmetry  is broken by marginal operators, the
renormalization group evolution of the difference
$\lambda_1'-\lambda_1$ takes it away from zero, even if it is chosen to
be zero at some arbitrary renormalization point~$\mu$. 

\begin{figure}[htbp]
\begin{center}
\includegraphics[width=0.4\textwidth]{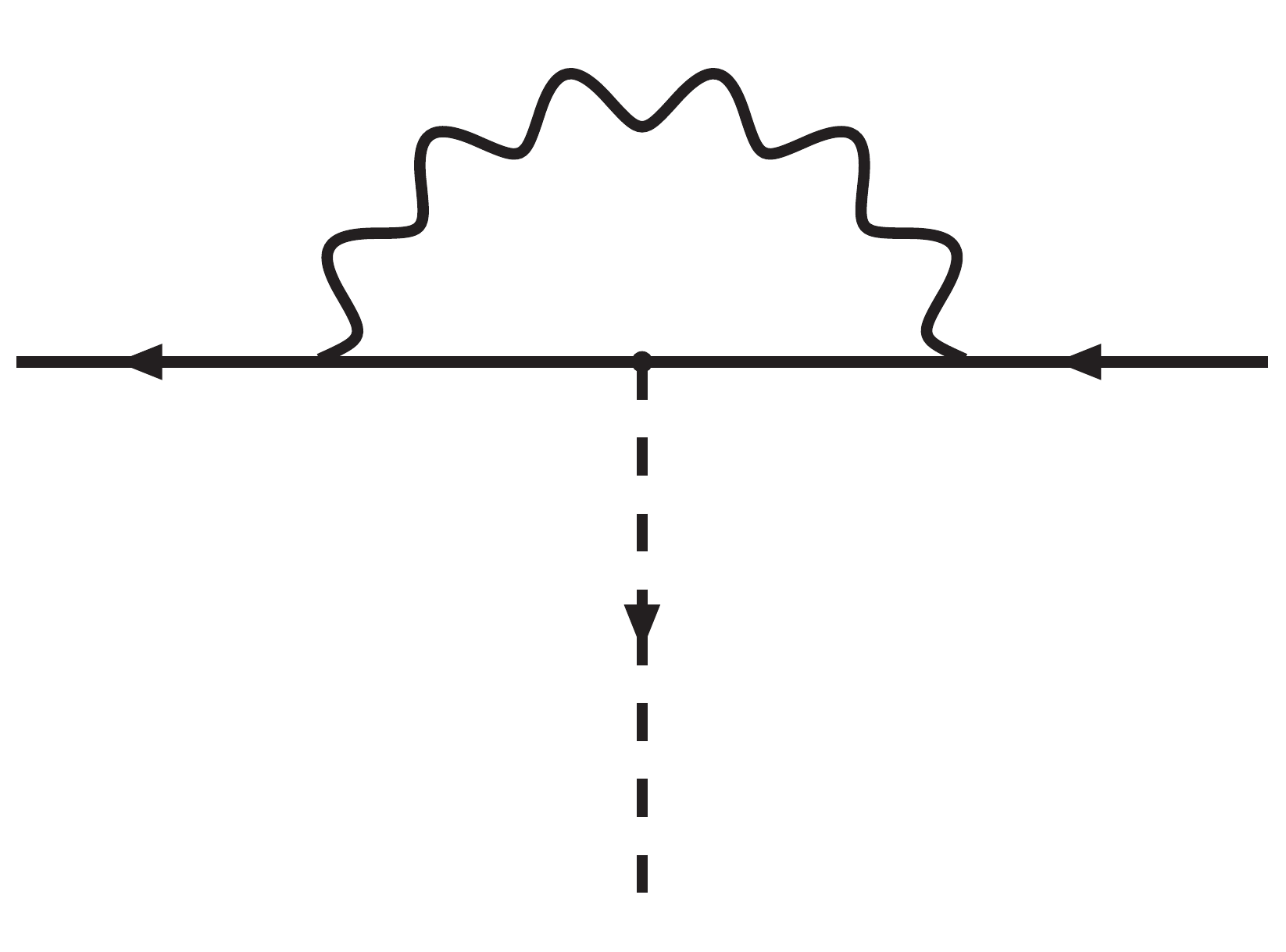}
\caption{Feynman diagram that contributes to the renormalization of
  the Yukawa couplings $\lambda_1$ and $\lambda_1'$. The wavy line
  represents a gauge boson of $U(1)_1$ and the solid and doted lines a
 spinor and a PGB, respectively.}
\label{fig:lambda-renor}
\end{center}
\end{figure}

As a check we have computed explicitly the one loop renormalization
group equations for these couplings (see Fig.~\ref{fig:lambda-renor}):
\begin{equation}
\mu\frac{\partial}{\partial\mu}\ln\left(\frac{\lambda_1}{\lambda'_1}\right)
=\left(\sfrac23-y\right)\frac{3g_1^{\prime2}}{16\pi^2}
\end{equation}
Here $y$ is the charge of $q_R$ under $U(1)_2$.  Details of the
calculation will be presented elsewhere\cite{us-in-prep}. If $\beta_{g'_1}=
(b/16\pi^2)g_1^{\prime3}$ then we can write the solution in terms of
the running coupling:
\begin{equation}
\label{eq:su3bkg}
\frac{\lambda_1(\mu)}{\lambda'_1(\mu)}
       =\frac{\lambda_1(\Lambda)}{\lambda'_1(\Lambda)}
\left(\frac{g'_1(\mu)}{g'_1(\Lambda)}\right)^{\frac{2-3y}{b}}
\end{equation}

The numerical value for $b$ can be obtained from the standard $QED$ beta function (see \cite{Machacek:1983tz})
\begin{equation}
	b = \frac{2}{3}\sum_{\text{Weyl fermion}}Y_{1i}^2 + \frac{1}{6}\sum_{\text{real scalar}} Y_{1i}^2
\end{equation}
To compute this, we need to introduce the Yukawa-type coupling for all
the other standard model quarks.  We will follow Perelstein
\cite{Schmaltz:2005ky} by noting that there is no need for
implementing collective symmetry breaking for the other standard model
quarks due to their small Yukawa couplings.  Thus the other ``up'' type
quarks Yukawa interaction can be introduced by
\begin{equation}
	-\lambda_\alpha^u f\bar q_{\alpha L}^{\PD i}\epsilon^{xy}  \Sigma_{ix}\Sigma_{3y}q_{\alpha R}^\PD
\end{equation}
where $\alpha=1,2$ is the quark family index. Similarly the other
``down'' type quark interactions can be introduced by
\begin{equation}
	-\lambda_\alpha^d f\bar q_{\alpha L}^{\PD i}\epsilon_{xy}  (\Sigma^*)^{ix}(\Sigma^*)^{3y}d_{\alpha R}^\PD
\end{equation}
here $\alpha=1,2,3.$ If we take $Y_2(q_R)=y$, then the $Y_1$ charge of
all the particles involved are
\begin{equation}
\label{eq:Y1Y2}
\begin{array}{c|ccccccc}
	 &q_{\alpha L} &q_{\alpha R} &d_{\alpha R} &u_L &u_R &H &\phi \\ \hline
	Y_1 &\sfrac{11}{30}-y &\sfrac23-y &\sfrac{1}{15}-y
        &\sfrac{13}{15}-y &\sfrac{13}{15}-y &1/4 &1/2\\
        Y_2 &y-\sfrac15 & y & y-\sfrac25 & y-\sfrac15 &y-\sfrac15 &1/4 &1/2
\end{array}
\end{equation}
Thus we get $b =\frac{1}{360} \left(2737-8832 y+10080 y^2\right)\ge
46/105$.  However, we note that the $y$ can be arbitrary.

We do not dwell on the numerics, since there are too many adjustable
parameters (the choice of $y$, the value of $U(1)$ couplings and
$\lambda_{1,2}(\Lambda)$ which however must satisfy \eqref{eq:bound1},
the value of the cutoff $\Lambda$). We simply note that $1/16\pi^2
\log(\Lambda/m_h)\sim1/16\pi^2 \log(100)\sim3\%$. Hence, even fine tuning
$\lambda_1(\Lambda)=\lambda'_1(\Lambda)$ generically produces a
difference $\lambda_1(m_h)-\lambda'_1(m_h)$ in excess of 1\%.

Note also that the same behavior must occur in the UV completion of the
\LLH\  model. After all, the terms in the Lagrangian  that break
$G_f$-symmetry  model the effects of symmetry breaking
interactions at short distances, that is, in the UV
completion.  The interactions in the UV completion that are responsible for the
quark Yukawa couplings cannot be taken to respect the $SU(3)$ symmetry
required for the collective symmetry argument. The breaking of the
$SU(3)$ symmetry in the UV completion is naturally much larger than
in \eqref{eq:su3bkg} since neither the $U(1)_1$ gauge coupling nor the
Yukawa couplings are asymptotically free.

\section{A no-go theorem}
\label{sec:no-go}
In this section we show the impossibility of constructing a theory
that implements without fine tuning the collective symmetry mechanism
on the terms responsible for quark and lepton masses. Let us begin by
stating in general terms what is required in order to implement the
collective symmetry mechanism.  Any given term in the Lagrangian has
to be symmetric under a subgroup $G_c$ of the flavor group $G_f$ under
which the higgs field transforms non-linearly, and in particular, with
a transformation that includes a constant shift.\footnote{Different
  terms in the Lagrangian may be invariant under different collective
  symmetry groups $G_c$.} In addition, there must not be any one loop
divergent radiative corrections that involve the coupling constants
for two different terms.

Of course there are additional requirements on each individual term in
the Lagrangian. In particular any one term must be invariant under
$G_w$, the gauged subgroup of $G_f$. We do not wish to specify this
gauge group, since one could look for realizations of the collective
symmetry mechanism in gauge groups other than the one of the
\LLH. Below we will only need to use the fact that this group
contains the electroweak gauge group, $G_{\text{ew}}=SU(2)\times
U(1)$, that this symmetry is linearly realized, {\it i.e.}, that
$G_{\text{ew}}\subset H$ so it remains unbroken at the scale at which
$G_f$ breaks to $H$, and that the higgs field must transform as a
doublet with hypercharge 1/2 under the electroweak group. 

The hidden fine tuning problem in the quark sector of the \LLH\
resulted from the fact that $G_c=SU(3)$ is not a symmetry of the
Yukawa term, because $G_c$ does not commute with $G_w$. The Yukawa
term in the Lagrangian is actually a sum of terms that are separately
invariant under the gauge group and the collection of terms can only
be symmetric under $G_c$ by fine tuning the separate coupling constants
at one scale. There are two ways that immediately come to mind in
which one could try to extend the \LLH\ model to get around this
problem. Either extend the gauge group so that $G_c$ itself is gauged
or obtain $G_c$ as an accidental symmetry. These, or other strategies
cannot work: below we will prove in generality that the collective
symmetry mechanism cannot work for terms other than the kinetic terms in
the Lagrangian.
 
\subsection{An $SU(7)/SO(7)$ example and its generalization to  $SU(N)/SO(N)$ }
It is simpler to understand the general case by first looking at an
explicit example. We can motivate this by the
following observation. If the $SU(3)$ collective symmetry that acts on
the first three rows and columns of $\Sigma$ is elevated to a gauge
symmetry, then the equality $\lambda_1'=\lambda_1$ is natural. Of
course, in the \LLH\ model this won't work because the $SU(3)$ is
broken at the scale $\Lambda$ at which $SU(5)$ breaks to $SO(5)$, and
the higgs is eaten at this scale. But perhaps one can construct a
theory based on a larger $G_f$ symmetry group with $SU(3)$ gauged and
the higgs still transforming non-linearly under some $G_c$ subgroup of
$G_f$. 

For example, one may consider a nonlinear sigma model based on
$G_f/H=U(7)/O(7)$ (with spinor fields in non-trivial representations
of the hyper-strong gauge group so that the $U(1)$ in
$U(7)=SU(7)\times U(1)$ is non-anomalous). Assume the $U(7)$ is broken
to $O(7)$ by a symmetric condensate, which transforms under $U(7)$ as
$\Sigma \rightarrow V\Sigma V^T$. Now gauge a $G_w=SU(3)\times
SU(2)\times U(1)^3$ subgroup of $U(7)$. The $SU(3)$ factor is
precisely the gauged version of the top-block collective symmetry group,
under which the royal triplet $\chi_L$ transforms as an actual
triplet. It is a straightforward, if lengthy, exercise to show that
the vacuum aligns correctly, that is, $G_w$ breaks to the electroweak
subgroup. One can identify $\Pi_{i4}$, and related entries, with the
higgs doublet. By suitably choosing the generators of the gauged
$U(1)^3$ symmetry one finds that the higgs field is the only light
PGB. 

Now introduce top quark couplings in a manner consistent with the
collective symmetry and without fine tuning of Yukawa couplings. Just
as in the \LLH\ model, in addition to the third generation quark
doublet $q_L$ and singlet $q_R$, introduce a pair of weak singlet Weyl
fermions $u_L$ and $u_R$ that transform as ${\bf{1}}_{1/6}$ under
$SU(2)_W\times U(1)_Y$. The singlet $u_L$ is combined with the doublet
$q_L$ into a triplet of the gauged $SU(3)$, precisely as in
\eqref{eq:triplet}. By suitably choosing the transformation properties
under the $U(1)^3$ we can ensure that the most general Yukawa
Lagrangian consistent with the symmetries, to lowest order in the
chiral expansion, is
\begin{equation}
  \label{eq:Ltop-full}
  {\cal L}_{\text{top}}= -f \lambda_1 \bar \chi_{Li}  (\Sigma^*)^{i4} q_R^\PD  
  -\frac12    f  \lambda_2 \bar  \chi_{Li}  
  \epsilon^{ijk}   \epsilon^{xy}    \Sigma_{jx}   \Sigma_{ky} u_R^\PD 
  +\hc
\end{equation}
where the indexes $i,j,k$ run over $1,2,3$ and 
$x,y$ over $5,6$. The problem with this model is that the $SU(3)$
symmetry does not protect the higgs. The collective symmetry required
is an $SU(4)$ acting on the top-left $4\times4$  block of
$\Sigma$. This in turn requires enlarging the true triplet to a
four-plet, which allows for more terms in the Lagrangian, which are
related by the $U(4)$ symmetry. However, this is not a good symmetry
of the Lagrangian and the added terms are related to the ones above only by 
imposing unnaturally a collective symmetry. This is precisely the same
problem we encountered with the \LLH. 

Let us generalize this to models with $SU(N)/SO(N)$ vacuum manifold,
parametrized by the $N\times N$ symmetric unitary matrix $\Sigma$. We
assume there is an $SU(2)\times U(1)$ gauged subgroup of
$SO(N)$. Without loss of generality we can take its embedding in $SU(N)$
as follows:
\begin{align}
\begin{split}
Q^a &= \frac{1}{2}
  \begin{pmatrix}
    \tau^{a} &0_{2\times (N-4)} &0_{2\times 2}\\
    0_{(N-4) \times 2} &0_{(N-4)\times (N-4)} &0_{(N-4)\times 2}\\ 
    0_{2\times 2} &0_{2\times (N-4)} & -\tau^{a*}
  \end{pmatrix}\\
  Y &= \frac12\text{diag}(1,1,y_3,\ldots, y_{N-2},-1,-1)
\end{split}
\end{align}
with $\sum y_i=0$. We assume further that the whatever other
interactions exist they align the vacuum along \eqref{eq:Sigma0} (with
the proper interpretation for the dimensions of the 0 blocks and the
center unit block). Then, as usual,
$\Sigma=\exp(i\Pi/f)\Sigma_{ew}\exp(i\Pi^T/f)=\exp(2i\Pi/f)\Sigma_{ew}$,
where in the last step we have chosen the broken generators to satisfy
$\Pi\Sigma_{ew}=\Sigma_{ew}\Pi^T$. The $N-4$ doublets  $\Pi_{ix}$ with $i=1,2$ and
$x=3,\ldots, N-2$, have hypercharge $1/2+y_x$. So any one of these for
which $y_x=0$ is a prospective higgs doublet. 

Under an infinitesimal $SU(N)$ transformation, $1+i\epsilon^aT^a$, the
matrix of goldstone bosons transforms as 
\begin{equation}
\label{eq:Pi-shift}
\delta\Pi=\frac{f}{2}(T^a+\Sigma_{ew}T^{aT}\Sigma_{ew}^\dagger)+\cdots
\end{equation}
where the ellipses stand for terms at least linear in the fields. We
are interested in finding a subgroup $G_c$ of $SU(N)$ under which the
higgs field transformation includes a constant shift. However any such
transformation does not commute with $SU(2)\times U(1)$.  Without loss
of generality we assume that the third entry has zero hypercharge,  $y_3=0$, so that
$\Pi_{i3}=\Pi_{3i}^*=\Pi_{(N-2)i}=\Pi_{i(N-2)}^*$ is the prospective
higgs doublet. Then $G_c$ must contain generators
\begin{equation}
  X=\begin{pmatrix}
    0_{2\times2}&x_{2\times1}&0_{2\times(N-3)}\\
    x^\dagger_{1\times2}&0_{1\times1}&0_{1\times(N-3)}\\
    0_{(N-3)\times2}&0_{(N-3)\times1}&0_{(N-3)\times(N-3)}
  \end{pmatrix}
 \end{equation}
or
\begin{equation}
  X=\begin{pmatrix}
    0_{(N-3)\times(N-3)}&0_{(N-3)\times1}& 0_{(N-3)\times2}\\
    0_{1\times(N-3)} &0_{1\times1}&x^{T}_{1\times2}\\
    0_{2\times(N-3)}&x^{ *}_{2\times1}&0_{2\times2}
  \end{pmatrix}
\end{equation}
with $x$ a complex two component vector. Both of these give the same
linear shift on the prospective higgs field, as can be verified by
computing $X+\Sigma_{ew}X^T\Sigma_{ew}^\dagger$.  It follows that for
either one of these generators we have
\begin{equation}
\label{eq:commut}
[Q^a,X]=X'
\end{equation}
where $X'$ is a generator of the form of $X$. This means that the $X$
generators transform under $SU(2)$ as a tensor operator; they are in
fact complex doublets with hypercharge 1/2, just like the higgs.  Now,
there are additional generators in $G_c$: at the very least it
contains the $SU(3)$ subgroup generated by the top-left or bottom
right $3\times3$ blocks. Together, $X$ and these additional generators
transform as a {\it reducible} representation of the electroweak
subgroup. It follows that a gauge invariant term in the Lagrangian
that is also invariant under $G_c$ is a sum of terms that are
individually gauge invariant. The only exception is when the term is
constructed of fields that are separately $SU(2)$ invariant, as is the
case of the $\lambda_2$ mass term, in \eqref{eq:Ltop}, in the \LLH\
model. But it is unnatural to choose the coefficients of these various
terms to make their sum $G_c$ invariant. This is because the gauge
interactions always break the symmetry. Gauge boson exchange Feynman
diagrams like that of Fig.~\ref{fig:lambda-renor} give divergent
corrections to these couplings, and the corrections do not preserve
the $G_c$ invariance.

We can relax one assumption above slightly. We do not need to assume the
vacuum alignment $\Sigma_{ew}$ is along \eqref{eq:Sigma0}. In order to
have a collective symmetry argument that one can already apply in the
gauge sector one needs the first and last two rows and columns to be
as in \eqref{eq:Sigma0}. But the central $(N-4)\times(N-4)$ block
does not have to be a diagonal matrix, only a unitary, symmetric
matrix. However, the argument goes through as before: the components
of $\Pi$ that we identify with the higgs are changed in precisely the
way that the shifts in \eqref{eq:Pi-shift} are modified and the rest
of the argument goes through unchanged.

The explicit proof for the case $G_f/H=SU(N)/Sp(N)$ is completely
analogous. 

\subsection{The general case}
We turn now to the general case. We assume that $G_w$ contains the
electroweak gauge group $G_{\text{ew}}=SU(2)\times U(1)$, with
$G_{ew}\subset H$. We further assume that a subset of goldstone bosons
can be identified with the higgs field. We consider a term in the
Lagrangian that is both symmetric under $G_{\text{ew}}$ and has a
collective symmetry $G_c$.  We show in the appendix that we only need to
consider semi-simple $G_c$, which we assume henceforth. 

That the higgs transforms linearly under the electroweak
gauge group means that there is a doublet $h$ in $\Pi$ that transforms
as
\begin{equation}
\delta_\epsilon h= i\epsilon^a\frac{\tau^a}2h+i\epsilon\frac12h
\end{equation}
under $SU(2)\times U(1)$. Under a group $G_c\in G_f$ 
$h$ transforms non-linearly,
\begin{equation}
\delta_\eta h= \eta^mx^m+\cdots
\end{equation}
where the implicit sum over $m$ is over all generators in $G_c$, for
some two component complex vectors $x^m$ and the ellipses stand for
terms at least linear in $h$.  One can redefine the basis of
generators in $G_c$ so that $x^m=0$ for $m\ge5$ and $x^m$ for
$m=1,\cdots, 4$ are unit vectors, with $m=1,3$ real and $m=2,4$
purely imaginary. Now consider the commutator,
\begin{equation}
(\delta_\eta\delta_\epsilon-\delta_\epsilon\delta_\eta)h= i\epsilon^a
\eta^m\frac{\tau^a}2x^m+i\epsilon\eta^m\frac12x^m+\cdots
\end{equation}
The commutator is again a non-linear transformation, a linear
combination of the same four generators in $G_c$ that shift the
higgs. In terms of the Lie algebra of $G_f$, denoting these generators
by  $X^i$, with\footnote{The index $i$ runs over 1,2 because the hermitian matrices
  break into a symmetric and an antisymmetric part, corresponding to
  the two real and two imaginary components of $x^m$, and also to 
  the real and imagnary components of  the higgs doublet.} $i=1,2$ and
the generators of $G_{\text{ew}}$ by $Q^a$ and $Y$, we read off
\begin{equation}
[Q^a,X^i]=\frac{i}2(\tau^a)^{ij}X^j,\qquad [Y,X^i]=\frac{i}2X^i
\end{equation}
This is precisely the statement in Eq.~\eqref{eq:commut}, derived there
from the explicit form of matrices, that the generators transform as
tensors of $G_{\text{ew}}$ with the same quantum numbers as the higgs
doublet, but we see now that it holds more generally, independently of
those explicit matrix representations. 

Since there is no semi-simple Lie algebra of rank 4, there must be
additional generators, and $[X^i,X^j]$ must give some of these
additional generators. Denote a non-vanishing commutator by $\hat
X^{ij}=[X^i,X^j]$. Using the Jacobi identity we see that
\begin{align}
  [Q^a,\hat X^{ij}]&=[Q^a,[X^i,X^j]]\\
  &=[X^i,[Q^a,X^j]]-[X^j,[Q^a,X^i]]\\
  &=\frac{i}2(\sigma^a)^{jk}X^{ik}-\frac{i}2(\sigma^a)^{ik}X^{jk}
\end{align}
So these generators also satisfy an equation like \eqref{eq:commut}
but transform in a representation in the tensor product of two
doublets. Continuing this way, considering commutators of the
generators we have so far, we can eventually generate the complete Lie
algebra and find that it breaks into sectors classified by irreducible
representations under $G_{\text{ew}}$.

We can use this to show that invariants under $G_c$ break into a sum
of terms separately invariant under $G_{\text{ew}}$. Any non-trivial
invariant must be a product of two combination of fields, one
transforming in some irreducible representation $R$ of $G_c$ and the
other as the complex conjugate $\bar R$. But from the previous
paragraph it follows that under $G_{\text{ew}}$ the representation $R$
breaks into a direct sum $R=r_1\oplus r_2\oplus\cdots$ of at least two
irreducible representations of $G_{\text{ew}}$. Therefore the product
$R\times \bar R$, contains the sum of at least two invariants under
$G_{\text{ew}}$, $r_1\times \bar r_1$ and $r_2\times \bar r_2$. Since
$G_c$ is not a symmetry of the theory (because the kinetic energy term
for the goldstone bosons is not invariant), the two (or more)
$G_{\text{ew}}$ invariants can be summed into a $G_c$ invariant only
by fine tuning coefficients in the Lagrangian. This completes the
argument.

It may not be self-evident that any non-trivial representation of
$G_c$ breaks into two or more representations under
$G_{\text{ew}}$. This can be shown by noting that the roots of the Lie
algebra, that is the weights of the adjoint representation, of $G_c$
break into a sum of irreducible representations of $G_{\text{ew}}$,
precisely the same representations that the generators fall
into.\footnote{This follows form considering the standard map
  $T^A\to|T^A\rangle$ of the generators of $G_f$, with $T^A|T^B\rangle
  = |[T^A,T^B]\rangle$. Then $Q^a|X^i\rangle =
  i/2(\sigma^a)^{ij}|X^j\rangle$ and so on for the other generators of
  $G_c$.} Then by following the same procedure as in establishing
branching rules for representations of Lie algebras, that is,
introducing projection operators in weight space, and using the fact
that the roots form irreducible representations, one obtains that
every representation of $G_c$ is decomposed into a sum of irreducible
representations of $G_{\text{ew}}$.

We remarked above that $U(1)$ factors in $G_c$ are ignored. This
requires some explanation. After all, one could conceivably take the
four broken generators to generate a collective symmetry group of
dimension 4, say $U(1)^4$ or $SU(2)\times U(1)$. But the $U(1)$
symmetries do not help insure the higgs remains massless. It is easy
to see why by considering first the familiar \LLH\ case. The
$\lambda_1$ and $\lambda_1'$ terms of \eqref{eq:Ltop-split} that need
to be related by collective symmetry to obtain necessary cancellations
in one loop graphs can be made separately invariant under several
$U(1)$ symmetries. In fact, the situation is reversed from the
semi-simple group case, where a representation $R$ of $G_c$ is a
direct sum of at least two irreducible representation of
$G_{\text{ew}}$. Since the irreducible representations of $U(1)$ are
one dimensional, it is $G_{\text{ew}}$ that relates several
irreducible representation of $U(1)$, and forces them together into a
term in the Lagrangian.

\section{Conclusions}
\label{sec:conc}
It is easy to see that the top quark couplings of the \LLH\ model are
fine tuned in order to preserve collective symmetry. Renormalization
effects break the symmetry. These effects must also be
present in the underlying UV completion so they cannot be dismissed as
small. In any case, they are generically too large for successful
phenomenology even purely in the context of the fine tuned \LLH\
model.

The problem cannot be circumvented by enlarging the model to one with a larger
underlying flavor symmetry group. Gauging collective
symmetry is not an option: it either gives a higgsless model or
again requires fine tuning to avoid  quadratically divergent
radiative corrections to the higgs mass. 

We have shown that the collective symmetry argument cannot be
implemented on the Yukawa couplings of little higgs models without
fine tuning. Of course, no-go theorems are only as good as its
assumptions. We did not prove that no model exists that can both
include top quarks and solve the little hierarchy problem. For
example, one can presumably partially supersymmetrize the model to
ensure the cancellation of top loop induced quadratic mass
divergences, at least at one loop. Still, in the absence of a novel
mechanism to suppress the quadratic divergences in the top quark
induced radiative corrections to the higgs mass without fine tuning,
much of the allure of these models is past.

\section*{Appendix}
We show that the four generators $X^i$ in $G_c$ that produce the
non-linear transformations of the four real components of the higgs
field are in a subalgebra that generates a semi-simple subgroup of
$G_c$. 

Our starting point are the commutation relations 
\begin{equation}
[Q^a,X^i]=R(Q^a)^{ij}X^j,\qquad [Y,X^i]=R(Y)^{ij}X^j.
\end{equation}
These are part of the algebra of $G_f$. We recall some basic facts
about compact Lie algebras (we follow and use the notation of
Ref.~\cite{georgi}). The Cartan subalgebra of $G_f$ is the largest set of
mutually commuting generators $H_i$,
$i=1,\ldots,r\equiv\text{rank}(G_f)$.  In the adjoint representation
define a vector space by the map $T^A\to|T^A\rangle$, where the $T^A$
are generators of $G_f$, and define the action of generators on this
vectors by $T^A|T^B\rangle = |[T^A,T^B]\rangle$. Moreover, define an
inner product on this space by $\langle
T^A|T^B\rangle=\text{Tr}(T^{A\dagger}T^B)$. Since the $H_i$ are
mutually commuting one can find a basis of the vector space
$H_i|E_\alpha\rangle=\alpha_i|E_\alpha\rangle$.  The states correspond
to the rest of the generators, $E_\alpha$. It follows that
$[H_i,E_{\alpha}]=\alpha_iE_{\alpha}$ and
$E_{-\alpha}=E^\dagger_{\alpha}$. Choose the generators of the Cartan
subalgebra to satisfy $\langle H_i|H_j\rangle=\delta_{ij}$.  It can be
shown
\begin{equation}
\label{eq:EEcomm}
[E_\alpha,E_{-\alpha}]=\sum_{i=1}^r\alpha_iH_i
\end{equation}
We intend to show that there is a basis of of Cartan generators for
which the four $X^i$ (or a linear combination of them) correspond to
two pairs $(E_\alpha,E_{-\alpha})$  that therefore do not commute among
themselves. 

We are free to take $H_1=Q^3$ and $H_2=Y$ as the first two members of
the Cartan subalgebra. Now, the standard representation 
\begin{align}
R(Y)=&\begin{pmatrix}-\tau^2/2&0\\ 0&-\tau^2/2\end{pmatrix},&
\qquad
R(Q^3)&=\begin{pmatrix}-\tau^2/2&0\\ 0&\tau^2/2\end{pmatrix},\\
R(Q^1)=&\begin{pmatrix}0&-\tau^2/2\\ -\tau^2/2&0\end{pmatrix},&
\qquad
R(Q^2)&=\begin{pmatrix}0&-i/2\,\mathbbm{1}_{2\times 2}\\ i/2\,\mathbbm{1}_{2\times 2}&0\end{pmatrix},
\end{align}
and we can make a transformation $X^i\to U^{ij}X^j$ to diagonalize
$R(Q^3)$ and $R(Y)$:
\begin{equation}
\begin{aligned}
\ [Y,X^1]&=-\frac12 X^1\\
[Y,X^2]&=+\frac12 X^2\\
[Y,X^3]&=-\frac12 X^3\\
[Y,X^4]&=+\frac12 X^4
\end{aligned}
\qquad
\begin{aligned}
\ [Q^3,X^1]&=-\frac12 X^1\\
[Q^3,X^2]&=+\frac12 X^2\\
[Q^3,X^3]&=+\frac12 X^3\\
[Q^3,X^4]&=-\frac12 X^4
\end{aligned}
\label{eq:comms4}
\end{equation}

The rest of the Cartan subalgebra can be chosen to commute with $X^i$,
as we now show. Suppose 
\begin{equation}
\begin{aligned}
\ [H_i,X^1]&=-\frac{a_i}2 X^1\\
[H_i,X^2]&=+\frac{a_i}2 X^2\\
[H_i,X^3]&=-\frac{b_i}2 X^3\\
[H_i,X^4]&=+\frac{b_i}2 X^4
\end{aligned}
\label{eq:commsH}
\end{equation}
Then the generators $H_i'=H_i-({a_i+b_i})/2\;Y-({a_i-b_i})/2\;Q^3$, commute
with $X^j$. 

Now, it is clear the the four $|X^i\rangle$ are among the states
$|E_{\alpha'}\rangle$ that satisfy
$H'_i|E_{\alpha'}\rangle=\alpha'_i|E_{\alpha'}\rangle$. Moreover the
vectors $\alpha'$ for $X^i$ are of the form
$(\pm\frac12,\pm\frac12,0,\ldots,0)$. Equation \eqref{eq:EEcomm} holds
only provided the $H_i$ that satisfy
$\text{Tr}(H_iH_j)=\delta^{ij}$. While the $H_i$ satisfy
this orthonormality condition, the new basis $H_i'$ generally does
not. Writing $H_i=V_{ij}H_j'$ gives an explicit set of eigenvectors of $H_i$,
\[
H_i|E_{\alpha'}\rangle=V_{ij}H'_j|E_{\alpha'}\rangle=V_{ij}\alpha'_j|E_{\alpha'}\rangle
\]
The eigenvectors are the same as those of $H_i'$, and hence the
$|X^j\rangle$ are still among them, but the eigenvalues have changed,
$\alpha_i=V_{ij}\alpha'_j$. But with this basis we can use
\eqref{eq:EEcomm}. Explicitly
\begin{align}
[X^1,X^2]&=-\frac12\sum_{i=1}^r(V_{i1}+V_{i2})H_i\\
[X^3,X^4]&=-\frac12\sum_{i=1}^r(-V_{i1}+V_{i2})H_i
\end{align}
We see that both commutators are non-vanishing, as we set out to demonstrate.

\vspace{2cm}
\begin{center}
  {{\bf{Acknowledgments}}}
\end{center}
Work supported in part by the US Department of Energy under contract
DE-FG03-97ER40546.

\end{document}